# Non-planar magnetoactive GES-based solar plasma stability

SOUVIK DAS and PRALAY KUMAR KARMAKAR*

Department of Physics, Tezpur University, Napaam, Tezpur 784028, India.
*Corresponding author. E-mail: pkk@tezu.ernet.in

**Abstract.** A laboratory plasma–wall interaction-based astrophysical gravito-electrostatic sheath (GES) model is methodologically applied to study the dynamic stability of the magnetoactive bi-fluidic solar plasma system in the presence of turbulence effect. The spherically symmetric GES-model formalism couples the solar interior plasma (SIP, internally self-gravitating, bounded) and the solar wind plasma (SWP, externally point-gravitating, unbounded) through the diffused solar surface boundary (SSB). A normal spherical mode ansatz results in a generalized linear quadratic dispersion relation depicting the modal fluctuations on both the SIP and SWP scales. A constructive numerical platform reveals the evolution of both dispersive and non-dispersive modal features of the modified-GES mode excitations. The reliability of the derived non-planar dispersion laws is concretized with the help of an exact analytic shape matching the previously reported results founded on the plane-wave approximation. It is found that the thermo-statistical GES stability depends mainly on the magnetic field, equilibrium plasma density and plasma temperature. It is speculated that the dispersive features are more pronounced in the self-gravitational domains against the electrostatic ones. The magneto-thermal interplay introduces decelerating (accelerating) and destabilizing (stabilizing) influences on the SIP (SWP), and so forth. At last, we briefly indicate the applicability of the proposed analysis to understand diverse helioseismic activities from the collective plasma dynamical viewpoint in accordance with the recent astronomical observational scenarios reported in the literature.

## 1. Introduction

The Sun, its circumambient atmosphere, interfacial involved scale-coupling and associated fluctuation dynamics form a long-standing challenging area of research yet to be well explored (Stix 1991; Dwivedi *et al*. 2007; Priest 2014; Goutam & Karmakar 2016; Gohain & Karmakar 2018). Originating through the hydromagnetic dynamo-action, the Sun's magnetic field plays as an energy source that affects all the intrinsic phenomena to define the self-confined solar plasma dynamics (Vazquez-Semadeni *et al.* 1998; Priest 2014). The most crucial problem in this thematic direction of helio-physics has hitherto been the solar wind and its consequent super(hyper)-sonic acceleration mechanism amid varied inhomogeneities excited by the diversified magnetoactive expansion effects (Vidotto 2021; Kellogg 2022).

On the laboratory plasma scales, the formation of the plasma sheath structure at the plasma boundary interface has attained a considerable amount of attention in recent years (Goutam & Karmakar 2016; Gohain & Karmakar 2018). This sheath, formed near the vicinity of the confining wall in laboratories is a thin non-neutral space charge layer (Gohain & Karmakar 2015, 2018). In the non-neutral lab-sheath formation processes, the Bohm ionic-flow criterion is required to be well fulfilled. According to this criterion, the ion flow speed at the sheath entrance zone must be at least as great as the ion sound phase speed (Gohain & Karmakar 2015). In the case of solar plasma, there is no explicit so-defined physical boundary wall located at any specified radial

This article is part of the Special Issue on "Waves, Instabilities and Structure Formation in Plasmas".

position except gravitational potential boundary. Applying the basic physics of plasma–wall interaction mechanism, as well known in laboratory plasmas (Goutam & Karmakar 2016; Gohain & Karmakar 2018), a theoretical model called gravito-electrostatic sheath (GES) model has been proposed (Dwivedi *et al*. 2007; Karmakar & Dwivedi 2011). According to this model, the entire solar plasma system is divided into two boundary scale regions – the bounded solar interior plasma (SIP) and the unbounded solar wind plasma (SWP). The SIP and SWP are coupled through a thin diffused concentric GES-based non-rigid solar surface boundary (SSB). It is noted that, unlike the laboratory-produced plasmas, the solar self-gravitationally confined plasmas are indeed radially inhomogeneous and non-uniform in nature due to inherent microphysical irregularities and non-uniformities in various forms (Karmakar *et al*. 2016). The solar self-gravitational wall strength has the maximum value at the SSB, as already well described by the gravitational Poisson formalism (Karmakar & Dwivedi 2011).

The GES model has been able to successfully explain several important problems related to the equilibrium (Dwivedi *et al*. 2007; Karmakar & Dwivedi 2011; Goutam & Karmakar 2016) as well as fluctuation (Karmakar *et al*. 2016; Gohain & Karmakar 2018) dynamics of the entire solar plasma system. Incorporating hydrodynamic turbulence as an important factor, a study of the equilibrium properties of the GES-based solar plasma in the presence of turbu-magnetic effects (arising from the joint action of turbulence and magnetic pressure) has recently been reported (Goutam & Karmakar 2016). It has been predicted therein that the solar GES structure gets modified due to the combined action of the magnetic field and acoustic turbulence. It has also been found that the GES width has now increased by 5.17% due to the turbu-magnetic effects against the original one (Dwivedi *et al*. 2007; Goutam & Karmakar 2016).

In the recent past, a semi-analytic classical non-relativistic study on the GES model fluctuation dynamics in the non-thermal $q$-non-extensive electron framework in the turbulent plasma background has been reported (Gohain & Karmakar 2018). A methodical normal radial mode analysis (planar) has reduced the entire fluctuating solar plasma model into a unique pair of linear generalized quadratic dispersion relations with a unique set of $K$-dependent coefficients on both the SIP and SWP scales. A quasi-linear coupling (via the non-linear GES-interplay action) between the gravitational (SIP) and acoustic (SWP) modes has also been reported to exist in the solar plasma system (Gohain & Karmakar 2018).

It is to be noted herewith that the above investigative works are indeed founded on the traditional quasi-classic approximation of short-wavelength perturbations (plane waves) (Karmakar *et al*. 2016) with no plasma polytropic assumption invoked. As a consequence, the said results need to be refined in light of the realistic dynamical factors remaining usually excluded for the sake of analytic simplicity. It eventually motivates us to carry out a continued GES-model study to analyze the non-thermal solar plasma fluctuation dynamics in the turbu-magnetized GES model fabric (Goutam & Karmakar 2016). It considers the entire solar plasma system in an amended spherical configuration as a polytropic magnetoactive plasma fluid in curved non-planar geometry. The formulation makes no quasi-classic plane-wave approximation of the conventional type.

In addition to the above, the excitation of instabilities in the open solar dynamics as a result of magnetoactive expansion effects has been reported very recently (Kellogg 2022). It is based on various observational data from the NASA-launched Parker Solar Probe (PSP), Coordinated Data Analysis Web (CDAWeb) and Solar Terrestrial Relations Observatory (STEREO). However, an adequate plasma-centric theoretical support for such current astronomical helioseismic observations is yet to be formulated.

The focal aim of this investigation, which is indeed motivated by the above factual scenarios here, is to see the magnetoactive GES-based solar plasma modal excitation and fluctuations dynamics; particularly, in the presence of magnetoactive effects on both the gravitational (bounded SIP) and acoustic (unbounded SWP) scales. We carry out a spherical normal mode analysis to investigate the active heliosiesmic stability features associated with the entire solar plasma system. The key novelty here lies in the formulation of the GES-based continued study in a thermal bi-fluidic model fabric. It properly considers the inclusion of polytropic effect, turbu-magnetic action, non-planar fluctuations with the relaxation of any kind of conventional quasi-classic approximation judiciously and so forth. It is interesting that, unlike the previous planar investigation developed on a simplified plane-wave approximation with non-thermal non-extensive electrons (Gohain & Karmakar 2018), the current non-planar GES fluctuations herein have shown a traditional radial coordinate ($\xi$) dependency. In parallel, it is also affected conjointly by the solar plasma density ($n_0$), magnetic field ($B_0$) and the solar core-to-electron temperature ratio ($T_0/T_e$) on both the SIP and SWP scales. It is also noticeable here that, the damping behavior of the waves in the acoustic regime (SWP) is more manifest than that of self-gravitational-regime (SIP).

The structural layout of the proposed manuscript is organized as per a standard format as follows. The physical model formalism is given in Section 2. The SIP and SWP calculation schemes are described in Sections 2.1 and 2.2, respectively. The results and discussions are portrayed in Section 3. The SIP and SWP scale analyses are distinctly illustrated in Sections 3.1 and 3.2, respectively. Finally, the main conclusions drawn from our analyses are summarily given in Section 4.

## 2. Physical model formalism

We consider a simplified magnetoactive solar plasma system consisting of the constitutive thermal electronic and ionic fluids coupled via the gravito-electrostatic Poisson formalism on the solar scales of space and time. The entire solar plasma system is embedded in a uniform magnetic field on both the constitutive bounded (SIP) and unbounded (SWP) scales via the interfacial SSB formed under the non-local action of the long-range gravito-electrostatic force balancing mechanism. The entire solar plasma system is mainly composed of ionized form of hydrogen (92%) and helium (8%). The relative abundance of the remaining elements, such as $\alpha$-particles, C, N, O and Fe is only about 0.01% (Stix 1991; Priest 2014). Hence, the presence of any other heavier constitutive species is summarily ignored to simplify the analysis so as to reveal the pure dynamical picture of the modal stability associated herewith. The fluid turbulence effects arising because of overlapping of multiple micro-kinematical scales of the constitutive particles is modeled with the help of the Larson logabarotropic equation of state (Vazquez-Semadeni *et al.* 1998). The influences sourced in any kind of non-ideal, gyro-fluidic and tidal actions are ignored. The initial distribution is presumed to uphold a hydrostatic homogeneous quasi-neutral equilibrium configuration on the spatiotemporal scales of current astronomical interest. The global quasi-neutral nature ($N_e \approx N_i = N$) of the entire solar plasma system as a direct outcome of the electrostatic Poisson formalism is justifiable on the realistic grounds that the asymptotic value of the Debye-to-Jeans scale length ratio is almost zero ($\lambda_{De}/\lambda_J \sim 10^{-20}$, (Dwivedi *et al.* 2007; Karmakar & Dwivedi 2011; Gohain & Karmakar 2015; Goutam & Karmakar 2016; Karmakar *et al.* 2016)). The application of the gravitational Poisson equation in our model analysis is reliably validated due to the fact that the Sun (SIP) formed via the *Jeansean* dynamic molecular cloud collapse (Stix 1991; Priest 2014) continues dynamically to remain as a self-gravitating fireball system at any time in nature through the so-called virial conditions for the initiation of the structure formation (Gohain & Karmakar 2018; Vidotto 2021). It hereby justifiably allows us to employ the *Jeansean* spatiotemporal scales as the standard astrophysical normalization scheme adopted for investigating the magnetoactive GES-induced fluctuation dynamics of current interest suitably. It is well known that, after reaching the main sequence stage (Stix 1991; Vidotto 2021), the Sun maintains a hydrostatically balanced state (gravito-thermal coupling). The outward randomized thermal (via contraction) and organized electrostatic forces (via plasma collective dynamics) jointly act against the radially inward self-gravity. This electro-thermal coupling hereby prevents the Sun from any further inward gravitational collapse. In other words, the self-gravitating subsonic SIP remains bounded under the action of the non-local self-gravitational influences (as non-uniform non-point source), unlike the unbounded supersonic SWP (as uniform point source). Hence, the solar self-gravitational effect is switched off on the SWP scale and transformed into a corresponding external gravity (Dwivedi *et al.* 2007; Goutam & Karmakar 2016; Gohain & Karmakar 2018). Hence, the gravitational Poisson equation becomes redundant on the SWP scale unlike the previous self-gravitating SIP scale.

It is well reported in the literature that the entire solar plasma behaviors obey a generalized polytropic equation of state enabling us to understand the corresponding plasma density and temperature fluctuations (Nicolaou *et al.* 2020). As a result, we apply a simple form of the polytropic equation of state for the effective plasma pressure, $p = K_p \rho^{\gamma} = K_p \rho^{(1+n_p^{-1})}$; where $K_p$ is the polytropic constant, $\gamma = 5/3$ is the polytropic exponent, and $n_p = (\gamma - 1)^{-1}$ is the corresponding polytropic index (Priest 2014; Goutam & Karmakar 2016). The active turbulence pressure considered here is given by a logabarotropic equation of state, $p_{turb} = p_0 \log(\rho/\rho_c)$, where $p_0$ is the mean (equilibrium) pressure, $\rho$ is the material density and $\rho_c$ is the heliospheric core material density (Vazquez-Semadeni *et al.* 1998). The polytropic, turbulence (logabarotropic) and magnetic pressures concurrently act on the complex solar plasma fluid system (Goutam & Karmakar 2016).

In order for our scale-free calculation ansatz to execute, we describe the adopted various standard symbolic significances associated with the well-known astrophysical normalization scheme (Dwivedi *et al*. 2007; Gohain & Karmakar 2015, 2018; Goutam & Karmakar 2016) as follows:

$m_{e(i)}$: Electron (ion) mass;
$T_0$: Solar core temperature (K);
$T_{e(i)}$: Electron (ion) temperature (K);
$\omega_J = c_s/\lambda_j = (4\pi n_0 m_i G)^{1/2} \approx 10^{-3}\,\mathrm{s}^{-1}$: Jeans frequency;
$a_0 = GM_\odot/c_s^2\lambda_j \approx 95$: External gravity-rescaling coefficient useful for temperature measurement;
$G = 6.67 \times 10^{-11}\ \mathrm{m}^3\ \mathrm{kg}^{-1}\ \mathrm{s}^{-2}$: Universal (Newtonian) gravitational coupling constant;
$k_B = 1.38 \times 10^{-23}\ \mathrm{m}^2\ \mathrm{kg}\ \mathrm{s}^{-2}\ \mathrm{K}^{-1}$: Boltzmann constant
$M_\odot = 2 \times 10^{30}$ kg: Mean solar mass;
$\alpha = B_0^2/\mu_0 n_0 k_B T_e = 2/\beta$: Magneto-thermal pressure coupling constant;
$\beta = 2\mu_0 n_0 k_B T_e/B_0^2$: Plasma-$\beta$ parameter (defined as the thermal-to-magnetic pressure ratio).

Adopting all the above customary notations relevant for the solar plasma dynamics (Dwivedi *et al*. 2007; Gohain & Karmakar 2015, 2018; Goutam & Karmakar 2016), we systematically describe our standard astrophysical normalization procedure in Table 1.

After following Table 1, we now proceed methodically to present the turbu-magnetoactive SIP and SWP calculation schemes in the normalized form separately. It is repeated that our aim is to employ a fluid dynamical treatment to both the constitutive electrons and ions actively coupled via the long-range gravito-electrostatic interplay (Poisson formalism) on both the subsonic SIP (bounded) and the supersonic SWP (unbounded) scales.

### 2.1 *SIP calculation scheme*

The self-gravitating SIP fluctuation dynamics is governed by the well-known set of the usual Jeans-normalized solar structuring equations in the above customary notations in the gravito-electrostatically coupling Poisson-closed form (Dwivedi *et al*. 2007; Goutam & Karmakar 2016). The same fundamental solar plasma equations in the normalized form are, respectively, presented as

$$\frac{\partial N_{e(i)}}{\partial \tau} + M_{e(i)}\frac{\partial N_{e(i)}}{\partial \xi} + N_{e(i)}\frac{\partial M_{e(i)}}{\partial \xi} + \left(\frac{2}{\xi}\right)N_{e(i)}M_{e(i)} = 0, \quad (1)$$

$$\frac{\partial M_{e(i)}}{\partial \tau} = s\frac{\partial \Phi}{\partial \xi} - \frac{1}{N_{e(i)}}\frac{\partial N_{e(i)}}{\partial \xi}\left[1 + \frac{1}{N_{e(i)}}\left(\frac{T_0}{T_e}\right)\right] - \frac{1}{N_{e(i)}}\alpha B_{az}^*\frac{\partial B_{az}^*}{\partial \xi} - \frac{\partial \Psi}{\partial \xi}, \quad (2)$$

$$B_{az}^*\left(\frac{\partial M_{e(i)}}{\partial \xi} + \frac{M_{e(i)}}{\xi}\right) + \frac{\partial B_{az}^*}{\partial \tau} = 0, \quad (3)$$

$$\frac{\partial^2 \Psi}{\partial \xi^2} + \left(\frac{2}{\xi}\right)\frac{\partial \Psi}{\partial \xi} = N_i, \quad (4)$$

$$\left(\frac{\lambda_D}{\lambda_J}\right)^2\left[\frac{\partial^2 \Phi}{\partial \xi^2} + \left(\frac{2}{\xi}\right)\frac{\partial \Phi}{\partial \xi}\right] = N_e - N_i = 0. \quad (5)$$

In the above, Equation (1) represents the electron (ion) continuity equation (Goutam & Karmakar 2016). Then, Equation (2) describes the electron (ion) momentum equation. Here, the electrostatic polarity phase factor is given as $s = +1$ (for electrons) and $s = -1$ (for ions). Also, Equation (3) represents the electron (ion) magnetic induction equation. Similarly, Equations (4) and (5) represent the gravitational and electrostatic Poisson equations designating the corresponding potential distributions sourced by the local density fields of matter and charge, respectively. Here, $(2/\xi)$ arises due to the consideration of spherical geometry ($\xi \nrightarrow \infty$); which would, otherwise, be absent for plane parallel geometry ($\xi \rightarrow \infty$). It is well known that solar plasmas are quasi-neutral in nature as already stated above. As a consequence, one finds $N_e \approx N_i = N$ as already mentioned above.

**Table 1.** Adopted astrophysical normalization scheme.

| Normalized parameter | Normalizing parameter | Typical value |
|---|---|---|
| Radial distance ($\xi = r/\lambda_j$) | Jeans length ($\lambda_j$) | $3.1 \times 10^8$ m |
| Time ($\tau = t/\omega_j^{-1}$) | Jeans time ($\omega_j^{-1}$) | $10^3$ s |
| Population density ($N_{e(i)} = n_{e(i)}/n_0$) | Mean SIP density ($n_0$) | $10^{30}\ \mathrm{m}^{-3}$ |
| Mach number ($M_{e(i)} = v_{e(i)}/c_s$) | Sound phase speed ($c_s$) | $3.09 \times 10^5\ \mathrm{m\ s}^{-1}$ |
| Gravitational potential ($\Psi = \psi/c_s^2$) | Sound phase speed squared ($c_s^2$) | $9.5 \times 10^{10}\ \mathrm{m}^2\ \mathrm{s}^{-2}$ |
| Electrostatic potential ($\Phi = e\phi/k_B T_e$) | Thermal potential ($k_B T_e/e$) | $10^2\ \mathrm{J\ C}^{-1}$ |
| Magnetic field ($B_{az}^* = B_{az}/B_0$) | Average SIP magnetic field ($B_0$) | $10^{-4}$ T |

We apply a spherical normal mode analysis (local treatment) over the defined hydrostatic homogeneous equilibrium configuration in a presumed spherical symmetry (with no polar and azimuthal counterparts) as

$$F(\xi,\tau) = F_0 + F_1(\xi,\tau), \tag{6}$$

$$F = \left[N_e\, N_i\, M_e\, M_i\, \Phi\, \Psi\, B^*_{az}\right]^T, \tag{7}$$

$$F_0 = [1\,1\,0\,0\,0\,0\,1]^T, \tag{8}$$

$$F_1 = \left[N_{e1}\, N_{i1}\, M_{e1}\, M_{i1}\, \Phi_1\, \Psi_1\, B^*_{az1}\right]^T. \tag{9}$$

Here, $F_0$ denotes the equilibrium values of the relevant parameters; while, $F_1(\xi,\tau) \sim (1/\xi)e^{-i(\Omega\tau - K\xi)}$ signifies the corresponding perturbations evolving as the restricted spherical waves with the Jeans-normalized angular frequency $\Omega$ and the Jeans-normalized angular wavenumber $K$. Applying Equation (6), we transform the coordination (direct) space $(\xi,\tau)$ into the Fourier (reciprocal) space $(K,\Omega)$. It transforms linear differential operators in the defined wave-space as $\partial/\partial\tau \rightarrow -i\Omega$, $\partial/\partial\xi \rightarrow (iK - 1/\xi)$ and $\partial^2/\partial\xi^2 \rightarrow (2/\xi^2 - K^2 - 2iK/\xi)$. Applying the above transformations in Equations (1)–(5) and equating the linearly fluctuating terms from both the sides, one finds the respective algebraic expressions in the $(K,\Omega)$-space as

$$M_{e1(i1)}\left(iK + \frac{1}{\xi}\right) - i\Omega N_{e1(i1)} = 0, \tag{10}$$

$$i\Omega M_{e1(i1)} = \left(iK - \frac{1}{\xi}\right)\left(s\Phi_1 + N_{e1(i1)} + \left(\frac{T_0}{T_e}\right)N_{e1(i1)} + \alpha B^*_{az1} + \Psi_1\right), \tag{11}$$

$$\Omega B^*_{az1} = M_{e1(i1)}K, \tag{12}$$

$$\Psi_1 = -\frac{N_{i1}}{K^2}, \tag{13}$$

$$N_{e1} = N_{i1}. \tag{14}$$

Applying the methodological procedure of elimination and simplification, Equations (10)–(14) decouple into a linear generalized quadratic dispersion relation for the SIP fluctuation dynamics cast as

$$\Omega^2 = \left(K^2 + \frac{1}{\xi^2}\right)\left[1 + \left(\frac{T_0}{T_e}\right) + \alpha K\left(K^2 + \frac{1}{\xi^2}\right)^{-1}\left(K + i\frac{1}{\xi}\right) - \frac{1}{K^2}\right]. \tag{15}$$

We use a composite frequency as $\Omega = (\Omega_r + i\Omega_i)$ in Equation (15) to obtain the real (normal and regular) and imaginary (perturbed and irregular) parts characterizing the GES instability modes triggered fundamentally by plasma flow currents, respectively, as

$$\Omega_r^2 - \Omega_i^2 = \left(K^2 + \frac{1}{\xi^2}\right)\left[1 + \left(\frac{T_0}{T_e}\right) + \alpha K^2\left(K^2 + \frac{1}{\xi^2}\right)^{-1} - \frac{1}{K^2}\right], \tag{16}$$

$$\Omega_r\Omega_i = \frac{\alpha}{2}\left(\frac{K}{\xi}\right). \tag{17}$$

Clearly, Equations (16) and (17) represent a pair of algebraic equations in the real and imaginary frequency parts $(\Omega_r, \Omega_i)$ in a mixed form. The solutions of these equations are, respectively, given as

$$\Omega_r(\xi,K) = \frac{1}{\sqrt{2}}\left[\chi + \sqrt{\chi^2 + \frac{\alpha^2K^2}{\xi^2}}\right]^{\frac{1}{2}}, \tag{18}$$

$$\Omega_i(\xi,K) = -\frac{1}{\sqrt{2}}\left[-\chi + \sqrt{\chi^2 + \frac{\alpha^2K^2}{\xi^2}}\right]^{\frac{1}{2}}, \tag{19}$$

where

$$\chi(\xi,K) = \left(K^2 + \frac{1}{\xi^2}\right)\left[1 + \left(\frac{T_0}{T_e}\right) + \alpha K^2\left(K^2 + \frac{1}{\xi^2}\right)^{-1} - \frac{1}{K^2}\right]. \tag{20}$$

It is obvious from Equations (18) and (19) that our proposed non-planar magnetoactive bi-fluidic GES stability features depend significantly on the solar core-to-electron temperature ratio $(T_0/T_e)$ and on the magneto-thermal pressure coupling constant $(\alpha)$. These are obviously the novel physical terms appearing in our proposed bi-fluidic spherical wave analysis against the pre-reported planar stability study (Gohain & Karmakar 2018).

The reliability of the derived non-planar SIP dispersion relation, Equation (15), is strengthened with the help of an exact analytic shape matching with the previously reported planar results founded basically on the plane-wave ($\xi \to \infty$) approximation (Gohain & Karmakar 2018).

### 2.2 *SWP calculation scheme*

On the unbounded SWP scale, the self-gravity action (dynamic variable) is switched off. The role of external gravity (Newtonian point source) comes into picture. As a result, the gravitational Poisson equation becomes superfluous (Dwivedi *et al.* 2007; Karmakar & Dwivedi 2011; Goutam & Karmakar 2016). Accordingly, the modified normalized momentum equations for the SWP constitutive species are presented in a generic form as

$$\frac{\partial M_{\mathrm{e(i)}}}{\partial \tau} = s\frac{\partial \Phi}{\partial \xi} - \frac{1}{N_{\mathrm{e(i)}}}\frac{\partial N_{\mathrm{e(i)}}}{\partial \xi}\left[1 + \left(\frac{T_0}{T_{\mathrm{e}}}\right)\frac{1}{N_{\mathrm{e(i)}}}\right] - \left(\frac{\alpha}{N_{\mathrm{e(i)}}}\right)B^*_{\mathrm{az}}\frac{\partial B^*_{\mathrm{az}}}{\partial \xi} - \frac{a_0}{\xi^2}. \quad (21)$$

The remaining equations (continuity, induction and electrostatic Poisson) describing the SWP dynamics are the same as derived earlier in the case of the SIP scale.

In order to model the SWP dynamics, we derive the same algebraic equations (as Equations (10), (12) and (14)) in the Fourier space $(K, \Omega)$ for the diverse perturbations from the normalized equations of continuity, magnetic induction and electrostatic Poisson, respectively. Now, the momentum equation (Equation 21) in the linearized form with the relevant perturbations reads as

$$\mathrm{i}\Omega M_{\mathrm{e1(i1)}} = \left(\mathrm{i}K - \frac{1}{\xi}\right)\left[s\Phi_1 + N_{\mathrm{e1(i1)}} + \left(\frac{T_0}{T_{\mathrm{e}}}\right)N_{\mathrm{e1(i1)}} + \alpha B^*_{\mathrm{az1}}\right] + \frac{2a_0}{\xi^2}N_{\mathrm{e1(i1)}}. \quad (22)$$

Employing the same method of elimination and decoupling over the normalized governing equations on the SWP scale, a linear generalized quadratic dispersion relation is procedurally derived and cast as

$$\Omega^2 = \left(K^2 + \frac{1}{\xi^2}\right)\left[1 + \left(\frac{T_0}{T_{\mathrm{e}}}\right) + \alpha K\left(K^2 + \frac{1}{\xi^2}\right)^{-1} \times \left(K + \mathrm{i}\frac{1}{\xi}\right)\right] - \frac{2a_0}{\xi^2}\left(\mathrm{i}K + \frac{1}{\xi}\right). \quad (23)$$

The real and imaginary parts of Equation (23) obtained with $\Omega = (\Omega_{\mathrm{r}} + \mathrm{i}\Omega_{\mathrm{i}})$ are, respectively, given as

$$\Omega_{\mathrm{r}}^2 - \Omega_{\mathrm{i}}^2 = \left(K^2 + \frac{1}{\xi^2}\right)\left[1 + \left(\frac{T_0}{T_{\mathrm{e}}}\right) + \alpha K^2\left(K^2 + \frac{1}{\xi^2}\right)^{-1}\right] - \frac{2a_0}{\xi^3}, \quad (24)$$

$$\Omega_{\mathrm{r}}\Omega_{\mathrm{i}} = \left(\frac{\alpha}{2} - \frac{a_0}{\xi}\right)\frac{K}{\xi}. \quad (25)$$

Applying the same root-finding techniques as in the SIP, we similarly obtain the respective roots $(\Omega_{\mathrm{r}}, \Omega_{\mathrm{i}})$ from Equations (24) and (25) presented as follows:

$$\Omega_{\mathrm{r}}(\xi, K) = \frac{1}{\sqrt{2}}\left[\zeta + \sqrt{\zeta^2 + \left(\alpha\frac{K}{\xi} - \frac{2a_0K}{\xi^2}\right)^2}\right]^{\frac{1}{2}}, \quad (26)$$

$$\Omega_{\mathrm{i}}(\xi, K) = -\frac{1}{\sqrt{2}}\left[-\zeta + \sqrt{\zeta^2 + \left(\alpha\frac{K}{\xi} - \frac{2a_0K}{\xi^2}\right)^2}\right]^{\frac{1}{2}}, \quad (27)$$

where

$$\zeta = \left[\left(K^2 + \frac{1}{\xi^2}\right)\left\{1 + \left(\frac{T_0}{T_{\mathrm{e}}}\right) + \alpha K^2\left(K^2 + \frac{1}{\xi^2}\right)^{-1}\right\} - \frac{2a_0}{\xi^3}\right]. \quad (28)$$

It is evident from Equations (26) and (27) that the magnetoactive GES-based SWP stability depends sensibly on the solar core-to-electron temperature ratio $(T_0/T_{\mathrm{e}})$ and magneto-thermal pressure coupling constant $(\alpha)$ in a similar fashion as previously found on the SIP scale from Equations (18) and (19) as well.

The validation of the derived non-planar SWP dispersion relation, Equation (23), is bolstered with the help of an exact analytic shape matching with the previously reported planar results grounded fundamentally on the plane-wave ($\xi \to \infty$) approximation (Gohain & Karmakar 2018).

## 3. Results and discussions

In this magnetoactive GES-based solar plasma stability analysis in the presence of bi-fluidic turbulence effects, a normal spherical modal procedure is employed to derive a pair of dispersion relations for the SIP (Equation 15) and SWP (Equation 23). The quasi-classic short-wavelength plane-wave approximation ($K\xi \gg 1$ or $\xi \gg \lambda$) is relaxed at the cost of the GES-centric spherical wave analysis for the first time. The quasi-linear relationship between the SIP and SWP scale fluctuations is structurally established through the coupling via the non-local long-range GES force-field action (Dwivedi *et al.* 2007; Karmakar & Dwivedi 2011; Goutam & Karmakar 2016; Karmakar *et al.* 2016; Gohain & Karmakar 2018). This quasi-linear auto-coupling is developed as the solar internal self-gravity (non-Newtonian) converts into an external gravity source (Newtonian) via the interfacial SSB. Apart from these obvious analytic aspects, both the dispersion relations derived for the SIP and SWP fluctuation dynamics are numerically analyzed in the relevant solar plasma system framework. Herein, we consider the solar core temperature and the electron temperature, respectively, as $T_0 = 10^7$ K and $T_e = 10^6$ K (Dwivedi *et al.* 2007; Karmakar & Dwivedi 2011; Goutam & Karmakar 2016; Karmakar *et al.* 2016; Gohain & Karmakar 2018).

### 3.1 *SIP scale description*

As shown in Figures 1 and 2, we depict, respectively, the color spectral profiles of the Jeans-normalized real frequency ($\Omega_r$) and imaginary frequency ($\Omega_i$) in a color-phase space. The phase space is defined with the variation in the Jeans-normalized radial coordinate ($\xi$) and the Jeans-normalized angular wavenumber ($K$) associated with the bounded SIP fluctuation dynamics (Equation 15) in two distinct respective $K$-regimes: $K$ = 0–1000 and $K$ = 0–100. We choose here the fixed input value of magneto-thermal pressure coupling constant as $\alpha = 1$. It is clearly seen that, at the vicinity of $K = 0$, the fluctuations show dispersive nature; while, as $K$ increases, we find a linear $\Omega_r = f(K)$ relationship (Figure 1). It depicts a non-dispersive acoustic-like behavior of the SIP waves. It is also observed that $\Omega_r$ remains almost constant with the $\xi$-variation (Figure 1). It signifies that the fluctuations have stable oscillatory propagation on the SIP scale.

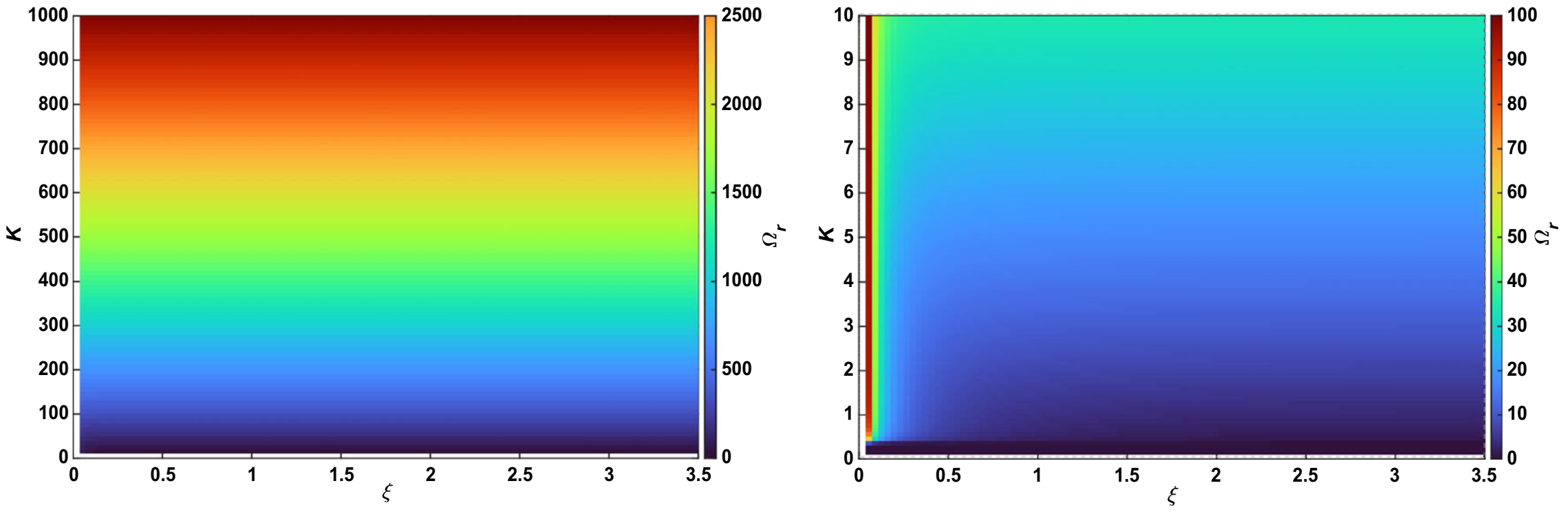

**Figure 1.** Color spectral profile of the Jeans-normalized real frequency ($\Omega_r$) with variation in the Jeans-normalized radial coordinate ($\xi$) and the Jeans-normalized angular wavenumber ($K$) associated with the SIP fluctuation dynamics. The different subplots correspond to different spectral scaling as shown.

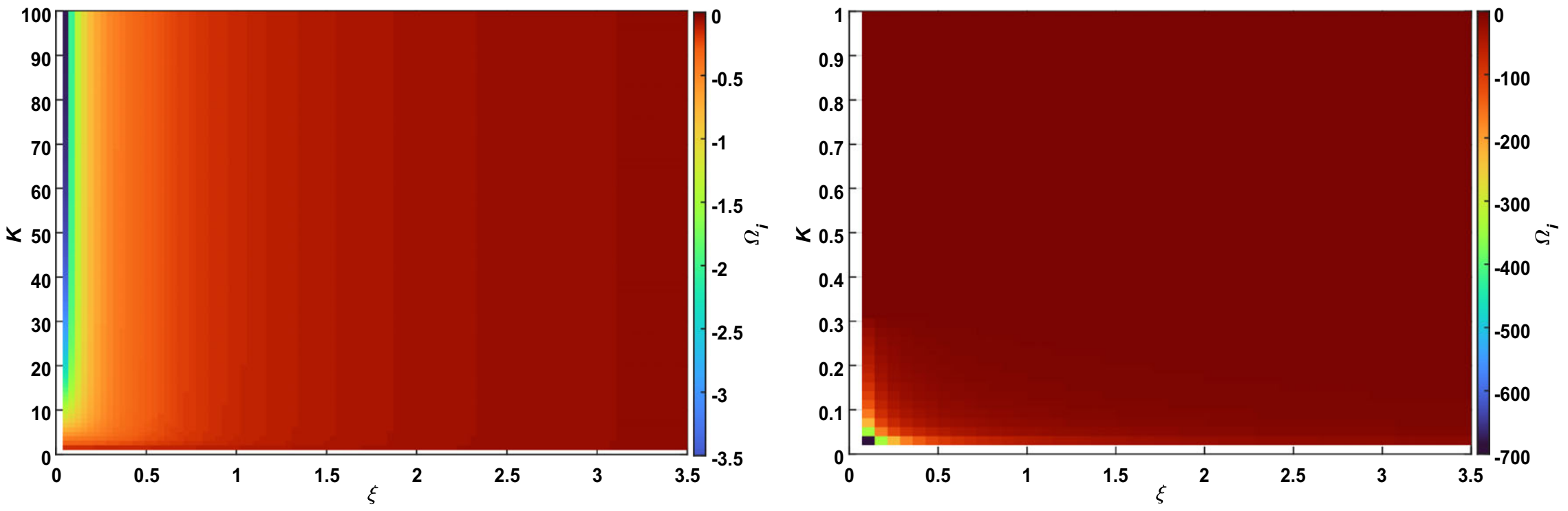

**Figure 2.** Color spectral profile of the Jeans-normalized imaginary frequency ($\Omega_i$) with variation in Jeans-normalized radial coordinate ($\xi$) and the Jeans-normalized angular wavenumber ($K$) associated with the SIP fluctuation dynamics. The different subplots correspond to different spectral scaling as shown.

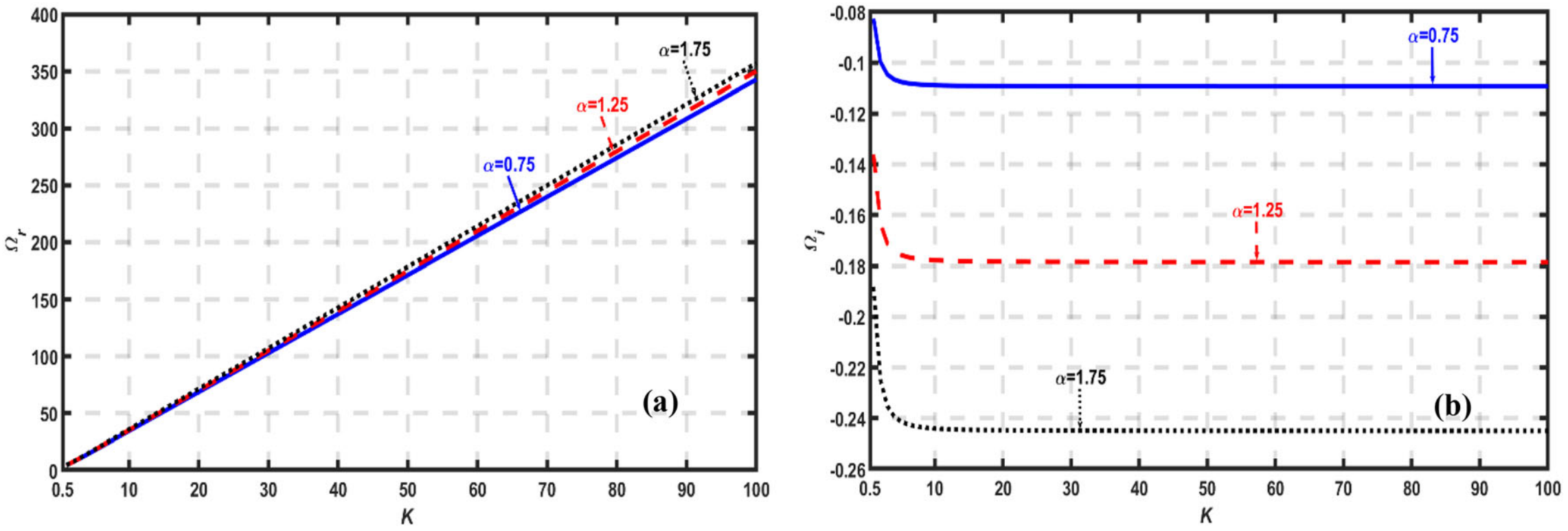

**Figure 3.** Profile of the Jeans-normalized (a) real ($\Omega_r$) and (b) imaginary ($\Omega_i$) frequency with variation in the Jeans-normalized angular wavenumber ($K$) for different values of the magneto-thermal pressure coupling constant ($\alpha$) associated with the SIP fluctuation dynamics.

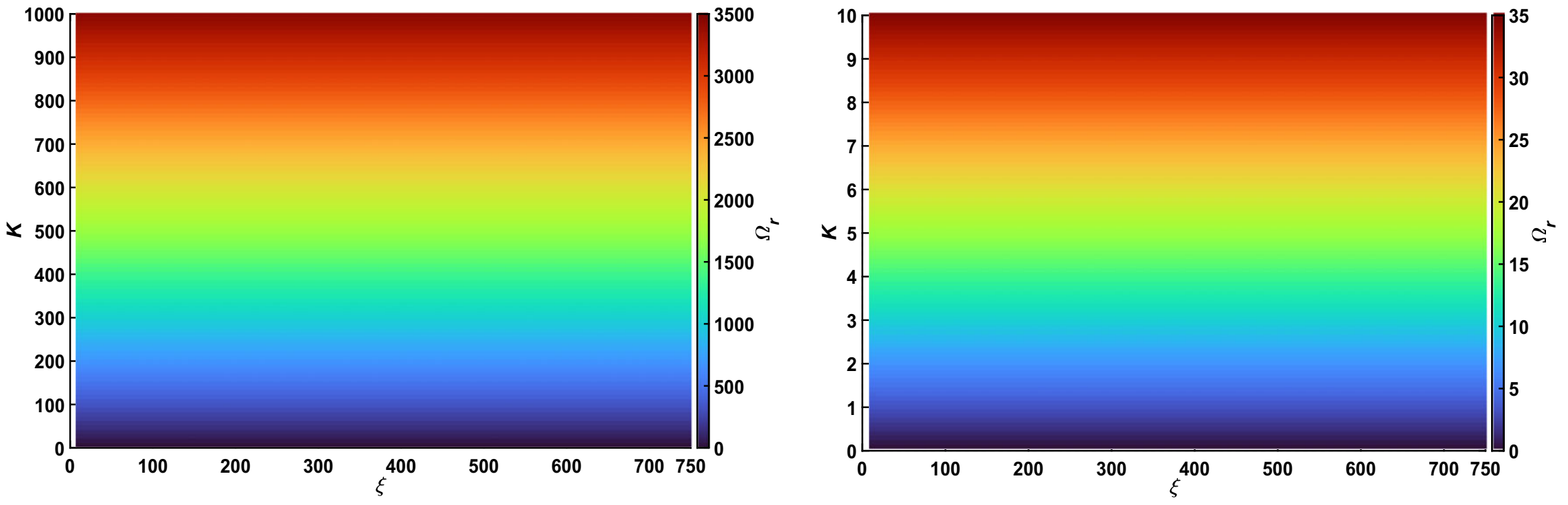

**Figure 4.** Color spectral profile of the Jeans-normalized real frequency ($\Omega_r$) with variation in the Jeans-normalized radial coordinate ($\xi$) and the Jeans-normalized angular wavenumber ($K$) associated with the SWP fluctuation dynamics. The different subplots correspond to different spectral scaling as clearly shown.

In contrast, $\Omega_i$ becomes almost constant in the $\xi$-space outwards against the center of the entire solar plasma mass distribution (Figure 2). It is, however, seen that in a very short range ($\xi \leq 0.2$), the fluctuations show slight damping behavior up to the vicinity of $K = 40$. Beyond it, the fluctuations show a propensity to stability (Figure 2). The entire picture is justifiably based on the complex long-range interplay of the non-local gravito-electrostatic origin.

As in Figure 3, we exhibit the profiles of (a) $\Omega_r$ and (b) $\Omega_i$ with variation in $K$ for different $\alpha$-values for the SIP fluctuation dynamics on the Jeans spatial scales ($\xi \sim 1$). It is obvious that, at the vicinity of $K = 0$, the perturbations show a dispersive nature. As the $K$-value increases, the collective fluctuations exhibit a non-dispersive type of behavior. It should be well representable with the help of a linear $\Omega_r = f(K)$ relationship (Figure 3a). It is also noticeable that $\Omega_r$ increases as $\alpha$ increases and vice versa. It could herewith be inferred that $\alpha$ is a decelerating agency here (Figure 3a). The overall helio-plasmic fluctuations show a sharp damping behavior up to the vicinity of $K = 5$ (Figure 3b). However, beyond it, $\Omega_i$ becomes almost constant (no damping interestingly). In fact, the $\Omega_i$-trend shows a sharp enhancing propensity as the $\alpha$-value increases actively and vice versa (Figure 3b) under the action of the non-local gravito-electrostatic interplay.

### 3.2 *SWP scale description*

In Figures 4 and 5, we depict the respective color spectral profiles of the Jeans-normalized real frequency part ($\Omega_r$) and imaginary frequency part ($\Omega_i$) with variation in the Jeans-normalized radial coordinate ($\xi$) and the Jeans-normalized angular wavenumber ($K$) of the unbounded SWP fluctuation dynamics with the same $K$-scaling as the SIP. Here, we take the standard value of the normalization coefficient as $a_0 = 95$. It is observed from Figure 4 that, on the unbounded scale, the perturbation waves behave purely acoustic-like without showing any strong dispersive nature. It implies that the effect of fluid turbulence pressure becomes too weak in the SWP regime to employ any observable effect in the wave propagation dynamics. There is no significant variation of $\Omega_r$ is observed in the $\xi$-space, as it is almost constant along $\xi$ (Figure 4). Thus, it represents the stable oscillatory propagation of the SWP fluctuations. Furthermore, the perturbation waves show a sharp damping behavior while going from the large wavelength scale to the shorter one (Figure 4). On the other hand, the variation of $\Omega_i$ along with $\xi$ (Figure 5) is similar to that observed in the SIP scale (Figure 2). It herewith signifies a stable oscillatory mode of the GES perturbation waves on both the bounded (SIP) and unbounded (SWP) spatiotemporal scales. Therefore, it can be inferred herewith that the helioseismic collective oscillation modes associated with the linear fluctuation dynamics of the solar plasmas are found to be scale-invariant under the quasi-hydrostatically balanced non-local GES-based force-field action.

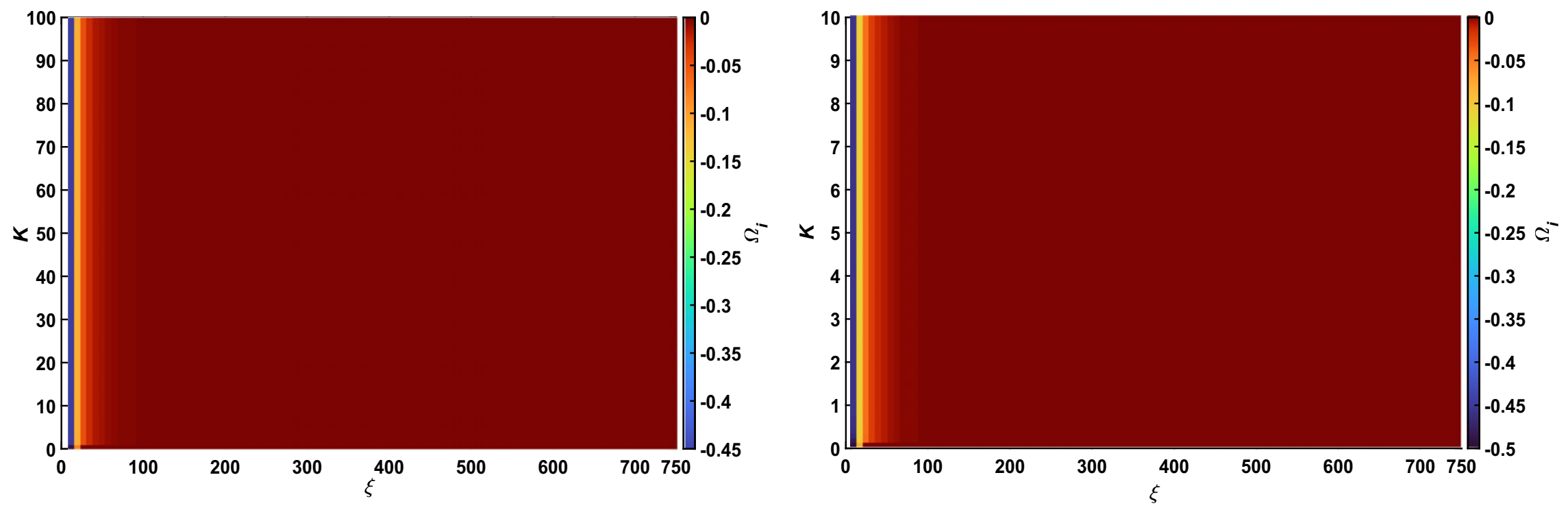

**Figure 5.** Color spectral profile of the Jeans-normalized imaginary frequency ($\Omega_i$) with variation in the Jeans-normalized radial coordinate ($\xi$) and the Jeans-normalized angular wavenumber ($K$) associated with the SWP fluctuation dynamics. The different subplots correspond to different spectral scaling as shown.

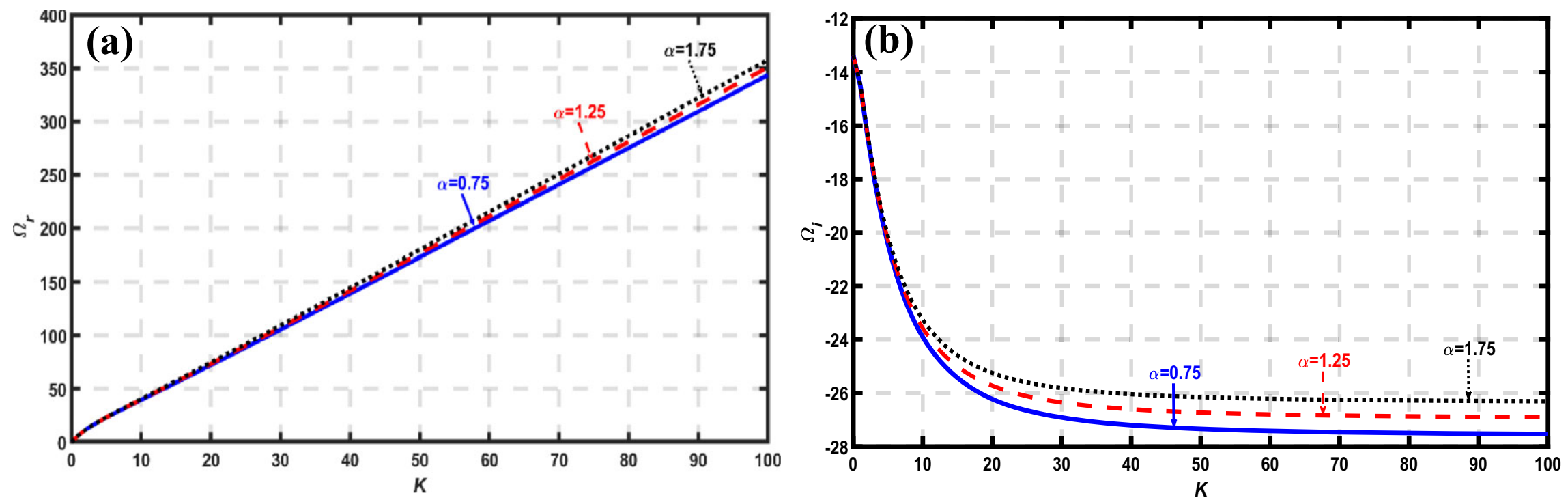

**Figure 6.** Profile of the Jeans-normalized (a) real ($\Omega_r$) and (b) imaginary ($\Omega_i$) frequency with variation in the Jeans-normalized angular wavenumber ($K$) for different values of the magneto-thermal pressure coupling constant ($\alpha$) associated with the SWP fluctuation dynamics.

Finally, Figure 6 portrays the profiles of (a) $\Omega_r$ and (b) $\Omega_i$ with variation in $K$ for different $\alpha$-values associated with the SWP fluctuation dynamics. It considers $a_0 = 95$ and $\xi \sim 1$ for a common spectral analysis. It is noticed that, up to the vicinity of $K = 5$, the SWP fluctuations show a dispersive nature. As $K$ increases, it becomes linearly dependent with $(\Omega_r = f(K))$, exhibiting purely a non-dispersive acoustic-like nature (Figure 6a). It is also observed that, in the SWP regime, $\Omega_r$ gains a growth trend as $\alpha$ increases (Figure 6a). It is similar to that found in the SIP regime (Figure 3a). Furthermore, it is also found that, the GES-based helioseismic fluctuations in the SWP scale show a sharp damping behavior up to the vicinity of $K = 15$, but beyond this $K$-domain, there is a slight damping nature (Figure 6b). Moreover, it is evidently seen that $\Omega_i$ shows a sharp decreasing nature as the $\alpha$-value increases and vice versa (Figure 6b). It indicates that $\alpha$ acts as a stabilizing factor to this fluctuation dynamics under the same physical background as above.

The new non-planar results presented above go in fair conformity with the previous planar predictions available in the literature in the same explorative GES-plasmic direction in a special case of plane parallel geometry (Gohain & Karmakar 2018). It hereby well ensures the reliability and validation of the proposed study of the solar plasma fluctuation dynamics at least in a qualitative manner. In a broader prospective glimpse, the main comparative features emanating from our methodological analysis in the light of the most relevant recent work founded on plane-wave analysis (Gohain & Karmakar 2018) could be pointwise summarized as follows:

1. In the previous planar instability work (Gohain & Karmakar 2018), the GES fluctuation dynamics has been studied in the framework of the non-thermal $q$-non-extensive electrons and isothermal ions in the framework of the basic plasma-wall interaction physical mechanism hitherto encounter in laboratory plasmas only. In our current model, we employ the polytropic equation of state for both the constitutive electrons and ions and study the GES-based instability analysis to characterize the turbu-magnetoactive helioseismic mode fluctuations in a new prospective view.
2. In the earlier planar work (Gohain & Karmakar 2018), the presence of the magnetic effects has been ignored; while, we consider the solar magnetic field effects herein to study the GES fluctuation characteristics in the fabric of the same plasma-wall interaction physics.
3. It is noteworthy that all the GES-based solar plasma stability descriptions are silent about the magneto-thermal coupling dynamics. It is shown herein that the magneto-thermal pressure coupling constant $(\alpha)$ acts as a decelerating (accelerating) and destabilizing (stabilizing) agency for the SIP (SWP) fluctuation dynamics (Figures 3b and 6b).
4. It is noticeable herein that, in the earlier planar work (Gohain & Karmakar 2018), a normal plane-wave analysis has been carried out; while, in our present study, the GES fluctuation dynamics is investigated with the help of a normal spherical wave analysis without any traditional quasi-classic approximation.
5. Lastly, in the extreme approximation of plane parallel geometry ($\xi \to \infty$, planar), instead of a spherical one ($\xi \nrightarrow \infty$, non-planar), our results well accord with the previous model predictions reported in the literature (Gohain & Karmakar 2018).

## 4. Conclusions

We herein report a detailed theoretic study of the fluctuation dynamics of the magnetoactive solar hydrodynamic system founded on the plasma-based turbu-modified GES model framework on the *Jeansean* spatiotemporal scales. It includes the combined action of plasma turbulence, geometrical curvature and magnetic pressure effects simultaneously. The helio-structure equations in exact analytic forms are constructed followed by a spherical perturbation mode analysis on both the interior (SIP) and exterior (SWP) solar plasma scales. It results in the modification of the GES-based solar plasma model by introducing a unique pair of generalized quadratic dispersion relations with a unique type of $K$-dependent modal behavior. It is shown that the developed magnetoactive linear GES stability not only depends on the radial position coordinate, but it is also affected by the solar core-to-electron temperature ratio $(T_0/T_e)$ and the magneto-thermal pressure coupling constant $(\alpha)$. Here, $\alpha$ plays as a decelerating (accelerating) destabilizing (stabilizing) source for the SIP (SWP) fluctuation dynamics, as shown in Figure 3(b) (Figure 6b). It is seen interestingly that the damping behavior of the GES-based fluctuations on both the SIP and SWP scales are more explicitly pronounced on the acoustic (SWP) scale (Figures 4 and 5) against the self-gravitational (SIP) scale (Figures 1 and 2).

It is pertinent to add further that, the numerical results, as presented in our non-planar analysis above (Figures 1–6), are purely based on a standard scale-free *Jeansean* calculation scheme to study the fluctuation dynamics associated with the magnetoactive non-planar solar plasma fluctuations founded on the GES model formalism. The quasi-linear coupling behavior of the gravitational ($K \to 0$) and the acoustic ($K \to \infty$) fluctuations investigated here fairly corroborates with those previously reported planar results available in the literature (Gohain & Karmakar 2018). Thus, it could hereby offer a validated reliability check-up of the GES-based solar plasmic fluctuation analysis paving the way for further realistic applicability in the emerging helioseismic direction of future space interest.

In the above context, it is widely seen that the entire solar plasma system is indeed anisotropic in nature with respect to the solar magnetic field axis (Sarfraz *et al*. 2022; Yoon *et al*. 2022). Both electron cyclotron instability (Sarfraz *et al*. 2022) and proton cyclotron instability (Yoon *et al*. 2022) significantly grow in such plasmas because of the respective excessive transverse temperature anisotropies. It indicates that a realistic refinement of the GES model with the inclusion of the velocity and temperature anisotropy of the plasma constitutive species is believed to be a significant problem for future investigation of the Sun and its circumvent atmosphere for the solar plasma community extensively.

We, finally, anticipate herewith that a better correlation and consistency is yet to bridge between the theoretical predictions developed on the proposed magnetoactive GES model formalism and the relevant observational solar data on various features of the solar plasma system as being recorded by the PSP and other Solar Orbiter Missions in the near future (Kasper *et al.* 2021; Vidotto 2021). In particular, one could herewith mention the natural excitation of diversified solar instabilities, which are driven by the magnetoactive expansion effects, reliably founded on various astronomical observational data from the NASA-operated PSP, CDAWeb and STEREO (Kellogg 2022).

## Acknowledgments

The authors acknowledge the active cooperation received from Tezpur University. The dynamic participation of colleagues of Astrophysical Plasma and Nonlinear Dynamics Research Laboratory, Department of Physics, Tezpur University, throughout is worth mentioning. The valuable contributions of the learned anonymous referees in the improvement of the original manuscript are duly noteworthy. The financial support received through the SERB Project, Government of India (Grant EMR/2017/003222), is duly recognized.